\documentstyle[preprint,aps,psfig]{revtex}
\tighten
\begin{document}
\draft
\preprint{\vbox{Submitted to Physical Review C}}
\title{Extracting the spectral function of ${}^{4}$He 
       from a relativistic plane-wave treatment}
\author{L.J. Abu-Raddad${}^{1}$
\footnote{Electronic address: laith@rcnp.osaka-u.ac.jp} 
and 
J. Piekarewicz${}^{2}$
\footnote{Electronic address: jorgep@csit.fsu.edu}}
\address{${}^{1}$ Theory Group, Research Center for Nuclear Physics,\\
Osaka University, 10-1 Mihogaoka, Ibaraki City, Osaka 567-0047, Japan}

\address{${}^{2}$Department of Physics, \\
         Florida State University, 
         Tallahassee, FL 32306, USA}
\date{\today}
\maketitle
 
\begin{abstract}
The spectral function of ${}^{4}$He is extracted from a plane-wave
approximation to the $(e,e'p)$ reaction using a fully relativistic 
formalism. We take advantage of both an algebraic ``trick'' and a 
general relativistic formalism for quasifree processes developed 
earlier to arrive at transparent, analytical expressions for all 
quasifree $(e,e'p)$ observables. An observable is identified for 
the clean and model-independent extraction of the spectral function. 
Our simple relativistic plane-wave calculations provide baseline 
predictions for the recently measured, but not yet fully analyzed, 
momentum distribution of ${}^{4}$He by the $A1$-collaboration from 
Mainz. Yet in spite of its simplicity, our approach predicts 
momentum distributions for ${}^{4}$He that rival some of the best 
nonrelativistic calculations to date. Finally, we highlight some 
of the challenges and opportunities that remain, both theoretically 
and experimentally, in the extraction of quasifree observables.
\end{abstract}

\narrowtext

\section{Introduction}
\label{sec:intro}
Electron scattering from nuclei is a common and powerful tool for
studying the structure of nuclei. The method relies on our superior
understanding of Quantum Electrodynamics (QED) and the relative ease
by which QED may be applied to a variety of processes, at least in the
one-photon-exchange approximation. In inclusive $(e,e')$ electron
scattering all nuclear-structure information is contained in two
dynamical quantities: the longitudinal and transverse response
functions. The longitudinal response is sensitive to the distribution
of charge in the nucleus while the transverse response samples the
distribution of currents and magnetization. Measurement of these
quantities in the quasielastic region is expected to be particularly
clean as the reactive content of the reaction is dominated by
quasifree proton knockout. If so, ``reduced'' longitudinal and
transverse response functions, obtained from the full nuclear
responses by dividing out the corresponding single-nucleon form
factor, should be equal. Yet a quenching of the longitudinal response
relative to the transverse one of 14\% in ${}^{4}$He and 50\% in
${}^{208}$Pb has been reported from a quasielastic $(e,e')$
electron-scattering measurement~\cite{zg94}. A similar, in fact even
larger 20-40\%, quenching in ${}^{4}$He has also been reported in the
semi-exclusive $(e,e'p)$ reaction at quasielastic
kinematics~\cite{du93}. In order to explain the
longitudinal/transverse (L/T) discrepancy a variety of explanations
have been proposed. These include medium-modifications to vacuum
polarization~\cite{hp89}, nucleon ``swelling~\cite{mu90}, and
Brown-Rho scaling~\cite{br89}. It is fair to say, however, that the
L/T anomaly remains an unsolved problem in nuclear physics.

The appeal of the $(e,e'p)$ reaction is due to the perceived 
sensitivity of the process to the nucleon momentum
distribution. Interest in this reaction has stimulated a tremendous
amount of experimental work at electron facilities such as NIKHEF, 
MIT/Bates, and Saclay, who have championed this effort for several 
decades. While it is undeniable that this reaction involves the best 
understood theory in all of physics (QED) many uncertainties remain 
due to the strongly-interacting character of the many-body system. 
It is hoped that with the advent of modern electron-scattering 
facilities, such as the Thomas Jefferson National Accelerator 
Facility (JLab) and Mainz, some of the remaining open questions will be 
answered. Indeed, in an attempt to elucidate the 
physics of the L/T anomaly discussed earlier, a systematic study of 
the longitudinal and transverse response functions from ${}^{3}$He 
and ${}^{4}$He is being conducted at the Mainz Microton (MAMI)
facility by the
A1-collaboration~\cite{Neu93,Flor98,Flor01,koz00,koz01a,koz01b}. 
Their extraction of ``experimental'' spectral functions and of
momentum distributions relies on a plane-wave-impulse-approximation 
(PWIA). In such an approximation the $(e,e'p)$ cross section is
proportional to the nucleon spectral function times an off-shell 
electron-proton cross section ($\sigma_{ep}$). Experimental analyses 
of this reaction employ, almost exclusively, the de~Forest's 
$cc1$ prescription for $\sigma_{ep}$ with both nucleon form factors 
unmodified from their free-space form~\cite{forest83}.

Stimulated by this new experimental thrust, we report here
relativistic plane-wave-impulse-approximation (RPWIA) calculations of
the $(e,e'p)$ cross section in the quasielastic region. Our motivation
for such a study is fourfold. First, we employ an established RPWIA
formalism, first introduced in Ref.~\cite{gp94} and recently extended
to the kaon-photoproduction reaction~\cite{abpi2000,thesis}, for the
study of the $(e,e'p)$ reaction in the quasielastic region. Second, we
use this formalism to compute the spectral function of ${}^{4}$He in
anticipation of the recently measured, but not yet fully analyzed,
$A1$-collaboration data from
Mainz~\cite{Flor98,Flor01,koz00,koz01a,koz01b}. Third, we take
advantage of the L/T separation at Mainz to introduce what we regard
as the cleanest physical observable from which to extract the nucleon
spectral function. Lastly, we highlight some of the challenges and
opportunities that remain in the calculation of quasifree observables.

There is a vast amount of literature on $(e,e'p)$ reaction in the 
quasifree region. Most relevant to our present discussion is the one 
pertaining to fully  relativistic 
calculations~\cite{Pick85,Pick89,Ud93,Ud96,Ud97,cdmu98,Ud98,Ud99,Ud01,Hed95,Joh96,Joh99,Meu01}.
An extensive set of these relativistic studies has been conducted by the 
{\it ``Spanish''} group of Udias and 
collaborators~\cite{Ud93,Ud96,Ud97,cdmu98,Ud98,Ud99,Ud01}. 
These studies have shown that the many subtleties intrinsic to the
relativistic approach challenge much of the ``conventional wisdom''
developed within the nonrelativistic framework and that, as a result,
a radical revision of ideas may be required. Relativistic effects
originating from medium modifications to the lower components of the
Dirac spinors and from the negative-energy part of the spectrum seem
to play an important role in the quasifree process. Indeed, the
much debated issue of short-range correlations at large missing
momenta~\cite{Weise72,Ste91,Lap93} can now be attributed, at least in
part, to contributions arising from the negative-energy
states~\cite{cdmu98,pr92}.

The power of the theoretical approach employed here lies in its
simplicity. Analytic expressions for the response of a mean-field
ground state may be provided in the plane-wave limit. The added 
computational demands placed on such a formalism, relative to that 
from a free on-shell proton, are minimal. The formalism owes its
simplicity to an algebraic trick, first introduced by Gardner 
and Piekarewicz~\cite{gp94}, that enables one to define a ``bound'' 
(in direct analogy to the free) nucleon propagator. Indeed, the 
Dirac structure of the bound nucleon propagator is identical to 
that of the free Feynman propagator. As a consequence, the power
of Feynman's trace techniques may be employed throughout the 
formalism.

The paper has been organized as follows. In Sec.~\ref{sec:formal}
some of the central concepts and ideas of the semi-exclusive $(e,e'p)$ 
reaction are reviewed. Special emphasis is placed on defining the
bound-state propagator and the simplifications that this entails
in the plane-wave limit. In Sec.~\ref{sec:results} we present our
results for ${}^{4}$He and discuss a (fairly) model-independent 
method for extracting the nucleon momentum distribution. Finally,
a summary and conclusions are presented in Sec.~\ref{sec:concl}.

\section{Formalism}
\label{sec:formal}

In Refs.~\cite{abpi2000,thesis} a general formalism has been developed
for the study of a variety of quasifree processes in the relativistic 
plane-wave impulse approximation (RPWIA). This formalism is now
applied to the  $(e,e'p)$ reaction in a mean-field approximation to 
the Walecka model~\cite{serwal86}. Although the use of a mean-field 
approach for a nucleus as small as ${}^{4}$He is questionable, we 
allow ourselves this freedom in order to establish a baseline against 
which more sophisticated approaches may be compared.

Following a standard procedure, an expression for the unpolarized 
differential cross section per target nucleon for the $(e,e'p)$
reaction is derived. We obtain,
 \begin{equation}
 \left(\frac{d^5\sigma }
  {dE_{e}^{\prime}d\Omega_{{\bf k}^{\prime}}
   d\Omega_{{\bf p}^{\prime}}}\right)_{\rm lab} = 
   \frac{4\alpha^{2}}{Q^{4}}
   \frac{|{\bf k}^{\prime}|} {|{\bf k}|} \mbox{ }
    {|{\bf p}^{\prime}|} \mbox{ } 
   {\big|{\cal M}\big|}^2 \;.
 \label{d5sigma}
\end{equation}
In the above expression ${\bf k}$, ${\bf k}^{\prime}$, and
${\bf p}^{\prime}$ denote the linear momentum of the incoming
electron, outgoing electron, and knocked-out proton, respectively.
The four-momentum transfer is defined in terms of the energy loss 
($\omega\!=\!E_{e}\!-\!E_{e}^{\prime}$) and the three-momentum 
transfer (${\bf q}\!=\!{\bf k}\!-\!{\bf k}^{\prime}$)
as $Q^{2}\!=\!{\bf q}^{2}-\omega^{2}$.
The transition matrix element ${\cal M}$ is given in a relativistic  
mean-field picture by
\begin{mathletters}
 \begin{eqnarray}
  \big|{\cal M}\big|^2 &=& l^{\mu\nu}\; W_{\mu\nu}\;, \\
   l^{\mu\nu} &=& \Big(
         k^{\prime\mu}k^{\nu} +
         k^{\mu}k^{\prime\nu} -
         g^{\mu\nu}(k\cdot k^{\prime}) \Big) \;, \\
   W^{\mu\nu} &=& \frac{1}{4(2j+1)}\sum_{s^\prime m} 
   \left(\overline{\cal U}({\bf p}^{\prime},s^{\prime}) 
   \,j^\mu\,{\cal U}_{\alpha m}({\bf p})\right)
   \left(\overline{\cal U}({\bf p}^{\prime},s^{\prime}) 
   \,j^\nu\,{\cal U}_{\alpha m}({\bf p})\right)^{*} \nonumber \\
   &=& \frac{1}{4}{\rm T}{\hbox{\lower 2pt\hbox{$r$}}}
   \Big(\left(\rlap/p^{\prime} + M \right)
   j^\mu  S_{\alpha}({\bf p})\,j^\nu \Big) \;.
  \end{eqnarray}
 \label{msquare}
\end{mathletters}
Here ${\cal U}({\bf p}^{\prime},s^{\prime})$ is the free Dirac
spinor for the knocked-out proton, normalized according to the
conventions of Bjorken and Drell~\cite{bj64}, while 
${\cal U}_{\alpha m}({\bf p})$ is the Fourier transform of the 
relativistic spinor for the bound proton. Note that $\alpha$
denotes the collection of all quantum numbers necessary to 
specify the single-particle orbital, except for the 
magnetic quantum number ($m$) which is indicated explicitly.
We have also introduced a ``bound-state propagator''
\begin{equation}
 S_{\alpha}({\bf p}) \equiv \frac{1}{2j+1}\sum_{m}
               {\cal U}_{\alpha m}({\bf p}) \,
      \overline{\cal U}_{\alpha m}({\bf p}) \;,
 \label{boundprop}
\end{equation}
normalized according to:
\begin{equation}
 \int\frac{d^{3}p}{(2\pi)^{3}}\,
  {\rm T}{\hbox{\lower 2pt\hbox{$r$}}}
  \Big(\gamma^{0} S_{\alpha}({\bf p})\Big)=
 \int\frac{d^{3}p}{(2\pi)^{3}}\,
 {\cal U}^{\dagger}_{\alpha m}({\bf p})\,
 {\cal U}_{\alpha m}({\bf p})=1 \;.
 \label{normalization}
\end{equation}
Here $j$ is the total angular momentum quantum number and 
$2j\!+\!1$ is the multiplicity of protons in the struck shell. 
It follows from simple kinematical arguments that the missing 
momentum ${\bf p}\!\equiv\!{\bf p}^{\prime}\!-\!{\bf q}$ is,
in a mean-field picture, identical to the momentum of the struck 
proton. It is the possibility of mapping the nucleon momentum 
distribution that makes the $(e,e'p)$ reaction so appealing.

We now invoke an algebraic trick first introduced in
Ref.~\cite{gp94} to simplify the expression for the
hadronic tensor $W^{\mu\nu}$. This technique is useful 
in quasifree processes as it enables one to cast the
bound-state propagator of Eq.~(\ref{boundprop}) into a 
form identical in structure to that of the free Feynman 
propagator. That is,
\begin{equation}
  S_{\alpha}({\bf p}) = 
 \Big({\rlap/{p}}_{\alpha} + M_{\alpha} \Big)\;,
 \label{salpha}
\end{equation}
where we have defined mass- and four-momentum-like
$\left[p_{\alpha}^{\mu}\!\equiv\!
(E_{\alpha},{\bf p}_{\alpha})\right]$ quantities according 
to
\begin{mathletters}
\begin{eqnarray}
  M_{\alpha} &=& \left({\pi \over p^{2}}\right)
                  \Big[g_{\alpha}^{2}(p) -
                       f_{\alpha}^{2}(p)\Big] \;, 
  \label{masslike}\\
  E_{\alpha} &=& \left({\pi \over p^{2}}\right)
                  \Big[g_{\alpha}^{2}(p) +
                       f_{\alpha}^{2}(p)\Big] \;,
  \label{epm} \\
  {\bf p}_{\alpha} &=& \left({\pi \over p^{2}}\right)
                   \Big[2 g_{\alpha}(p) 
                          f_{\alpha}(p)\hat{\bf p} 
                   \Big]  \;.
\end{eqnarray}
\end{mathletters}
Moreover, they satisfy the ``on-shell relation''
\begin{equation}
  p_{\alpha}^{2}=E_{\alpha}^{2}-{\bf p}_{\alpha}^{2}
                =M_{\alpha}^{2} \;.
 \label{onshell}
\end{equation}
In these expressions $g_{\alpha}(p)$ and $f_{\alpha}(p)$ are 
the Fourier transforms of the upper and lower components of 
the bound-state Dirac spinor, respectively~\cite{gp94}.
Using this form of the bound-state propagator the hadronic
tensor simplifies to:
\begin{equation}
   W^{\mu\nu} = \frac{1}{4}
    {\rm T}{\hbox{\lower 2pt\hbox{$r$}}}
    \Big(\left(\rlap/p^{\prime} + M \right)
    j^\mu 
    \left(\rlap/p_{\alpha}+M_{\alpha}\right) 
    j^\nu \Big) \;.
 \label{hadronic}
\end{equation}

The obvious similarity in structure between the free and 
bound propagators results in an enormous simplification: 
powerful trace techniques developed elsewhere may now be 
employed here to compute all $(e,e'p)$ observables. 
Although the focus of this paper is the unpolarized cross 
section [Eq.~(\ref{d5sigma})] the formalism may be extended 
without difficulty to the case in which the electron, the 
outgoing proton, or both, are polarized. Yet, in order to 
automate this straightforward but lengthy procedure, we rely 
on the {\it FeynCalc 1.0}\cite{mh92} package with 
{\it Mathematica 2.0} to calculate all the necessary traces. 
For a general electromagnetic current operator for the proton, 
the output from these symbolic manipulations is transparent 
enough so that the sensitivity of the cross section to the 
various quantities in the problem may be assessed. Indeed,
such a simplification will prove useful later in identifying 
the optimal observable from which to extract the spectral 
function. It is important to note, however, that this enormous 
simplification would have been lost had distortions been 
included in the formalism. Even so, the plane-wave approach 
discussed here, and used in most experimental extractions 
of the spectral function, is qualitatively useful. Moreover, 
if the main effect of distortions is to induce an overall 
suppression of the cross section without affecting significantly 
the distribution of strength, the plane-wave formalism provides 
solid quantitative predictions for a variety of spin 
observables~\cite{abpi2000,thesis}.

Yet an important open question remains: what constitutes a 
suitable form for the nucleon electromagnetic current? A 
ubiquitous form given in the literature is  
\begin{equation}
  j^\mu (q) = F_1(q^2) \gamma^\mu + i F_2(q^2) 
             \,\sigma^{\mu\nu} {q_{\nu} \over {2 M}}
\label{cc2}\;.
\end{equation}    
While this form is certainly general, as only two form factors 
are required to fully specify the electromagnetic current for an 
on-shell nucleon, the form is not unique. Indeed, many other 
forms --- all of them equivalent on-shell --- may be used. For 
example, through a Gordon decomposition of the current one 
arrives at
\begin{equation}
  j^\mu (q) = (F_1 + F_2) \gamma^\mu - F_2 
              \frac{(p^\prime + p)^\mu}{2M} 
\label{cc1}\;.
\end{equation}    
However, as soon as one of the nucleons goes off its mass shell, an
off-shell choice must be made. This decision is crucial, as various 
on-shell equivalent choices may yield vastly different results. This 
off-shell ambiguity remains one of the most serious obstacles in the 
field. Several attempts have been made in the literature to overcome 
this hurdle. Perhaps the most celebrated treatment is due to 
de~Forest who uses physical constraints, such as current conservation, 
to reduce this ambiguity~\cite{forest83}. He imposes this condition 
on the two forms of the electromagnetic current given above
[Eqs.~(\ref{cc2})~and~\ref{cc1}] and produces what are known in the
literature as the $cc2$ and the $cc1$ forms, respectively. Although 
noteworthy, this  effort does not resolve the ambiguity. For example, 
there is no unique way to impose current conservation; one may 
eliminate either the time component or the longitudinal component 
of the three-vector current~\cite{cdmu98}. Alternatively, one may 
adopt some guiding principle, such as vector-meson-dominance, to go 
off the mass shell. Here we adopt the ``natural'' choice by simply 
extrapolating off the mass shell the $cc2$-form, without imposing 
further constraints on the single-nucleon current.

\section{Results}
\label{sec:results}

\begin{figure}
\bigskip
\centerline{
  \psfig{figure=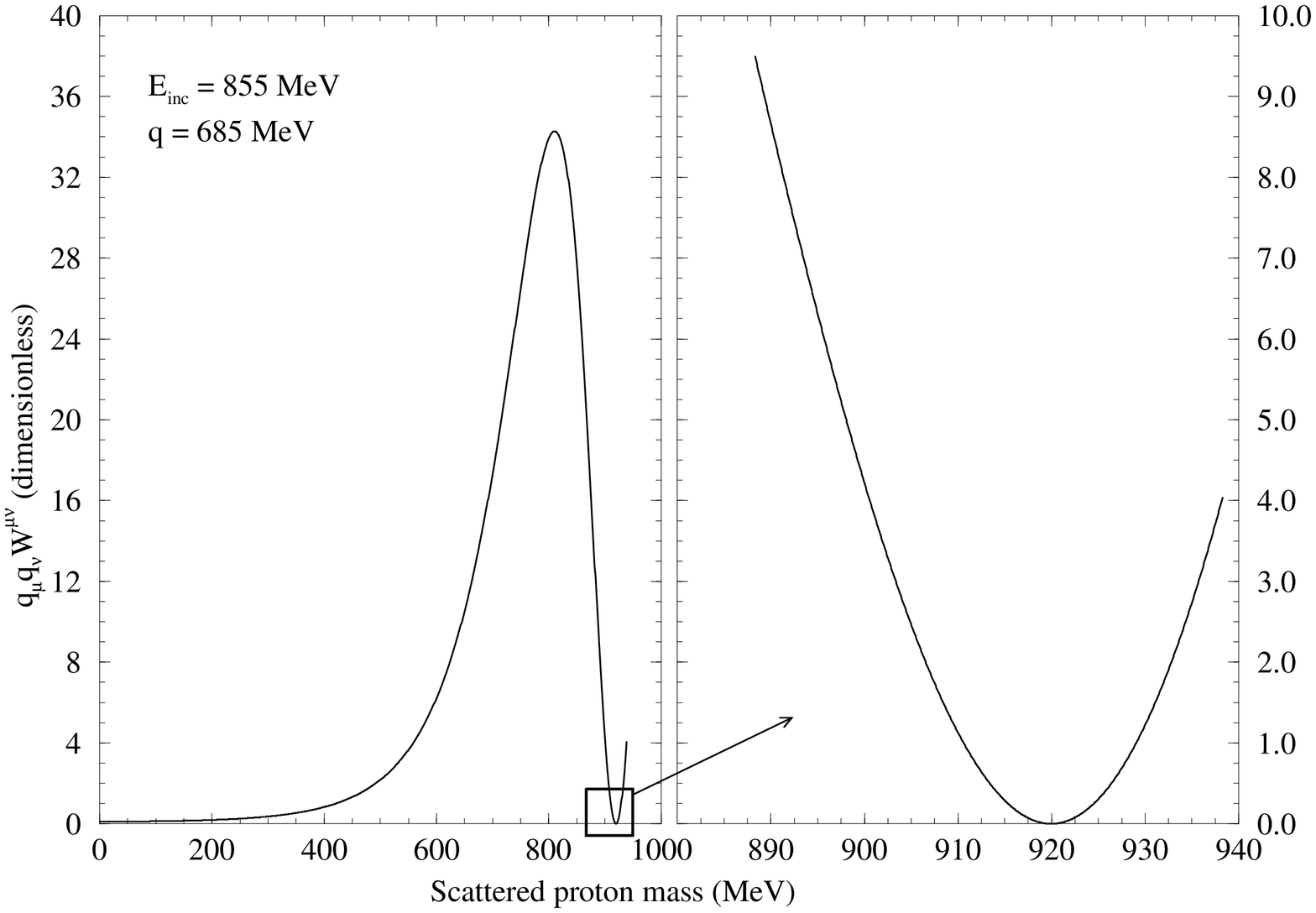,height=3.5in,width=5.5in,angle=-90}}
 \vskip 0.1in
 \caption{The gauge variance term $q_\mu q_\nu W^{\mu\nu}$ for
	  ${}^{4}$He as a function of the scattered proton mass
	  calculated in parallel kinematics for an incident photon
	  energy of $E_{\rm inc}=855$~MeV and a momentum transfer of
	  $q=685$~MeV. The right panel is a magnification of the boxed
	  area in the left panel.}
 \label{fig1}
\end{figure}

As de~Forest has done in the past, we now attempt to impose some
approximate form of gauge invariance. Yet rather than concentrating
on the nucleon current, we focus directly on the nuclear responses.
First, however, we address some important issues in this regard. For
any mean-field treatment of the $(e,e'p)$ reaction to be gauge
invariant, the mean-field potential for the bound proton must be
identical to the distorting potential for the emitted proton. This
represents a challenging task. Indeed, mean-field approximations to
the nuclear ground state give rise to real, local and
energy-independent potentials that are in contradiction to the complex
and energy-dependent potentials that are needed to describe the
propagations of the outgoing proton. Thus, present-day calculations of
$(e,e'p)$ observables are presented with a dilemma.  Calculations that
use the same (real and energy-independent) mean fields to generate
both the bound single-particle wave-function and the distorted wave
satisfy gauge invariance but miss some of the important physics, such
as absorption, that is known to be present in the outgoing channel. On
the other hand, calculations that incorporate the correct physics via
a phenomenological optical potential are known to violate current
conservation~\cite{Ud01}.  We offer here no solution to this complicated
problem. Rather, we impose gauge invariance ``ad-hoc'' by adjusting
the effective nucleon mass of the emitted proton so that the
``gauge-variance'' term, $q_\mu q_\nu W^{\mu\nu}$, be minimized. This
procedure, with perhaps its unexpected outcome, is displayed in
Fig.~\ref{fig1}.  It shows that by decreasing the proton mass by about
$20$~MeV, one can restore gauge invariance in the calculation: $q_\mu
q_\nu W^{\mu\nu}\!=\!0$.  Although by no means fundamental, this
``poor-man'' distortion ensures the conservation of gauge invariance
without compromising the clarity of the formalism.

The essence of the experimental extraction of the spectral function 
is based on a nonrelativistic plane-wave result~\cite{Fru84}:
\begin{equation}
S(E , {\bf p}) = {1\over p^\prime E_{p}^{\prime}\sigma_{eN}} \;
\frac{d^6\sigma}  
  {d{E^{\prime}_e d\Omega_{{\bf k}^{\prime}}
  {d{E}^{\prime}_p d\Omega_{{\bf p}^{\prime}}}}} \;.
 \label{speca} 
\end{equation}
However, this procedure is problematic. First, the quasifree cross
section [the numerator in Eq.~(\ref{speca})] suffers from the
off-shell ambiguity; different on-shell equivalent forms for the 
single-nucleon current yield different results. Second, the problem 
gets compounded by the use of an elementary electron-proton cross 
section ($\sigma_{eN}$) evaluated at off-shell
kinematics~\cite{forest83}. Finally, the projection of the bound-state 
wave-function into the negative-energy sector as well as other
relativistic effects spoil the assumed
factorization of the cross section derived in the nonrelativistic 
limit~\cite{cdmu98}.

Insights into the role of relativistic corrections, particularly those 
concerned with negative-energy states, may be gained by introducing
the completeness relation in terms of free (plane-wave) spinors:
\begin{equation}
  \sum_s  
     \left[{\cal U}({\bf p},s)\, 
  \overline{\cal U}({\bf p},s) -
           {\cal V}({\bf p},s) 
  \overline{\cal V}({\bf p},s)\right] = 1\;.
\label{comp}
\end{equation}    
Naively, one would expect that the projection of a positive-energy
bound state into a negative-energy plane-wave state would be 
vanishingly small. This, however, it is not the case~\cite{pr92}. 
At the very least one must recognize that the positive-energy 
plane-wave states, by themselves, are not complete. Moreover, it 
has been shown that the projection of the bound-state spinors 
into the negative-energy states dominate at large missing momenta 
and may mimic effects perceived as ``exotic'' from the nonrelativistic 
point of view, such as an asymmetry in the missing-momentum 
distribution~\cite{gp94} or short-range correlations~\cite{pr92}. 
Indeed, Caballero and collaborators have confirmed that these 
contributions can have significant effect on various observables, 
especially at large missing momenta~\cite{cdmu98}.  

\begin{figure}
\bigskip
\centerline{
  \psfig{figure=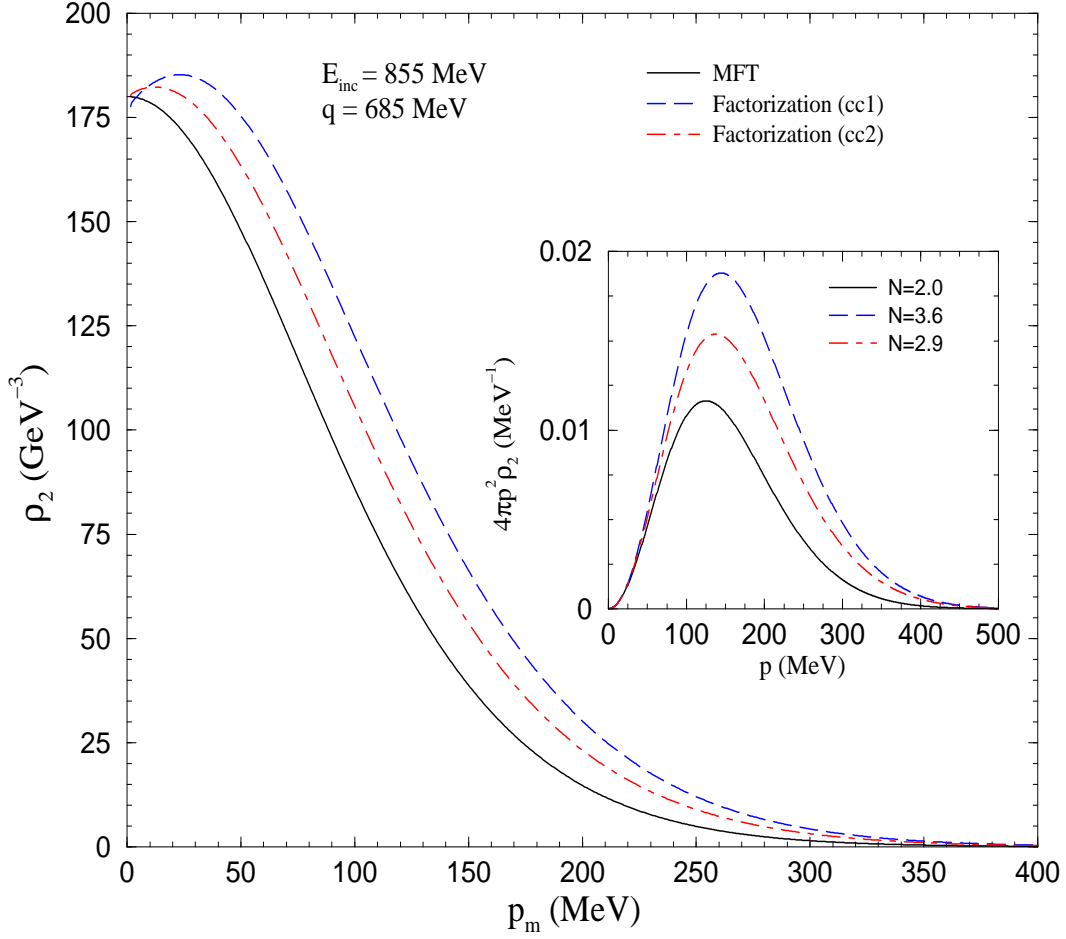,height=5.0in,width=5.5in,angle=-90}}
 \vskip 0.1in
 \caption{The proton momentum distribution $\rho_2$ for ${}^{4}$He as
	  a function of the missing momentum calculated at an incident
	  photon energy of $E_{\rm inc}=855$~MeV and a momentum
	  transfer of $q=685$~MeV. The solid line is the relativistic
	  mean-field calculation, while the dashed and dot-dashed
	  lines display the momentum distribution extracted from a
	  factorization approximation using the $cc1$ and $cc2$
	  prescriptions for $\sigma_{eN}$, respectively. The inset
	  shows the corresponding integrands from which the shell
	  occupancy may be extracted.}
 \label{fig2}
\end{figure}

To ``resolve'' the off-shell ambiguity it has become ubiquitous in the
field to use the de~Forest $cc1$ prescription for evaluating the
elementary cross section $\sigma_{eN}$ --- irrespective of the form of
the electromagnetic current adopted to compute the quasifree cross
section. This is the standard procedure used in comparing theoretical
calculations of the spectral function to experiment. We may elect here
to conform to tradition and use the de~Forest $cc1$ prescription to
compute $\sigma_{eN}$ in Eq.~(\ref{speca}), but at a cost. A price
must be paid because of the inconsistency in using one prescription
for evaluating the single-nucleon current $\sigma_{eN}$ and a
different one ($cc2$) to evaluate the quasifree cross section.  To
illustrate this point we display in Fig. \ref{fig2} the proton
momentum distribution defined by
\begin{equation}
 \rho_2({\bf p}) = \int S(E,{\bf p})\,dE \;.
\label{rho}
\end{equation}    
Note that the subscript ``2'' in $\rho_2$ stands for two-body breakup.
The graph displays the ``canonical'' momentum distribution (solid line)
obtained from the Fourier transform of the $1S^{1/2}$ proton
wave-function [see Eq.~(\ref{epm})]. Note that this canonical momentum
distribution has been normalized, as it is done experimentally, to the
total number of protons in the shell (2 for the case ${}^{4}$He).  The
other two curves were extracted from the quasifree cross section by
adopting either the de~Forest $cc1$ choice for $\sigma_{eN}$ (dashed
line) or the $cc2$ prescription (dot-dashed line). In both cases the
quasifree cross section has been computed using the ``vector-tensor''
form of the electromagnetic current, as given in Eq.~(\ref{cc2}). The
inset on the graph shows the integrand from which the occupancy of the
shell may be computed. It is evident that the conventional $cc1$
prescription of de~Forest greatly overestimates $\rho_2$ (it
integrates to 3.6). We attribute this deficiency to the lack of
consistency: the quasifree cross section has been evaluated using the
$cc2$ form of the current while the elementary amplitude uses the
$cc1$ form. One can improve the situation by adopting the $cc2$ form
in the evaluation of both. Yet significant differences remain; while
the off-shell ambiguity has been reduced, it has not been fully
eliminated. Moreover, the factorization assumption is only
approximate, as it neglects the projection of the relativistic wave
function onto the negative-energy spectrum and other relativistic effects.

\begin{figure}
\bigskip
\centerline{
  \psfig{figure=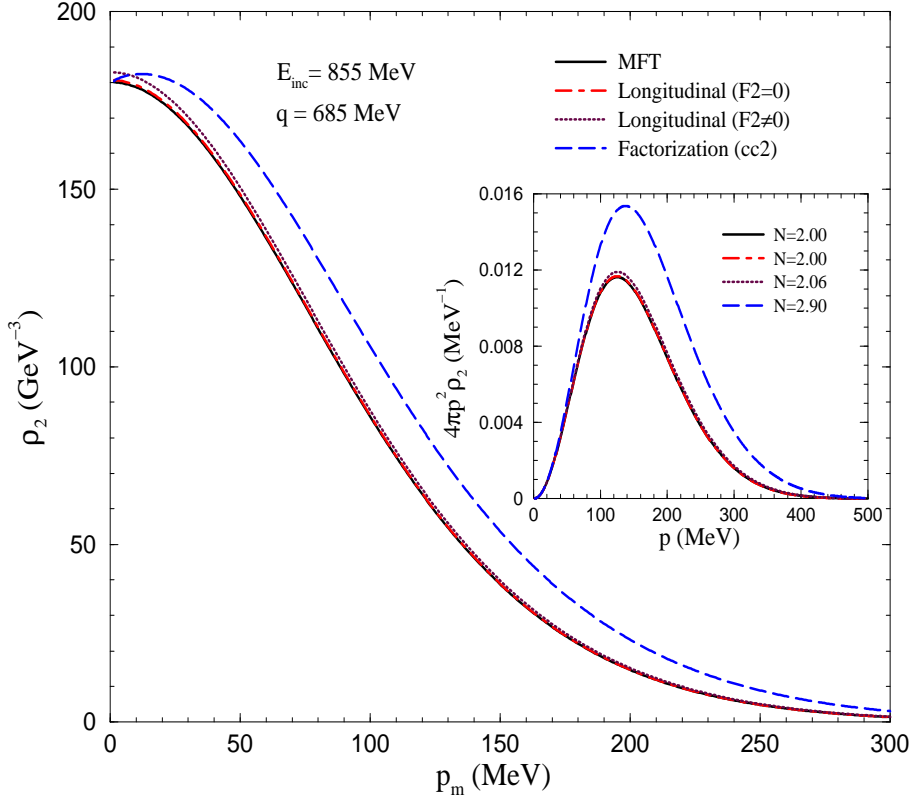,height=5.0in,width=5.5in,angle=-90}} 
 \vskip 0.1in 
 \caption{The proton momentum distribution $\rho_2$ for ${}^{4}$He as
	  a function of the missing momentum calculated at an incident
	  photon energy of $E_{\rm inc}=855$~MeV and a momentum
	  transfer of $q=685$~MeV. The solid line is the relativistic
	  mean-field calculation, while the dashed and dot-dashed
	  lines display the momentum distribution extracted from the
	  longitudinal response $R_L$ without including (dot-dashed)
	  and including (dotted) the contribution from the anomalous
	  form factor $F_{2}$. Finally, the dashed curve is obtained
	  by using the factorization approximation with the $cc2$
	  prescription for $\sigma_{eN}$. The inset shows the
	  corresponding integrands from which the shell occupancy may
	  be extracted.}
 \label{fig3}
\end{figure}

While a consistent relativistic treatment seems to have spoiled the
factorization picture obtained from a nonrelativistic analysis, and
with it the simple relation between the cross-section ratio and the 
spectral function [Eq.~(\ref{speca})], the situation is not without
remedy. Having evaluated all matrix elements of the electromagnetic 
current analytically in the plane-wave limit, the source of the
problem can be readily identified. Upon evaluating the coincidence 
cross section, one learns that the off-shell ambiguity is manifested 
in the form of several ambiguous ``kinematical'' factors. For example, 
one must decide what value to use for the energy of the struck proton.
Should it be the binding-energy of the struck proton or should it be 
the on-shell value? This is not an easy question to answer. Energy
conservation demands that the energy be equal to the binding energy 
($E_{\rm bin}\!=\!E_{p}^{\prime}\!-\!\omega$) yet the equivalence 
between the various forms of the electromagnetic current is derived 
assuming the on-shell dispersion relation 
($E_p\!=\!\sqrt{{\bf p}^2+M^2}$). 
This is 
one of the many manifestations of the off-shell ambiguity: kinematical 
terms that are well defined for on-shell spinors become ambiguous 
off-shell. In Ref.~\cite{forest83} de Forest resolves the ambiguity, 
by fiat, using the on-shell choice. Perhaps a better option may be 
looking for an observable, that even though might be more difficult 
to isolate experimentally, it may display a weaker off-shell 
dependence than the unpolarized cross section. To do so we examine 
the various components of the hadronic tensor. We find, perhaps not 
surprisingly, that the longitudinal component of the hadronic tensor 
could be such a model-independent observable. Ignoring  (for now) 
the anomalous part of the electromagnetic current, the Dirac-Dirac 
component of the longitudinal tensor [see Eq.~(\ref{hadronic})]
becomes:
\begin{equation}
  W^{00}_{\rm DD} = 
  F_1^2 \left[ M_{\alpha}M - p_{\alpha} \cdot p^\prime 
               + 2 E_{\alpha} E_{p}^\prime \right] =
  F_1^2 \left[ M_{\alpha}M + E_{\alpha} E_{p}^\prime 
               + {\bf p}_{\alpha}\cdot{\bf p}^\prime \right]\;. 
\label{w00dd}
\end{equation}    
This expression depends exclusively on $p_{\alpha}$ and $p^\prime$, 
which are unambiguous.
Note that for scattering from a free on-shell nucleon the above
expression becomes:
\begin{equation}
 W^{00}_{\rm DD} \mathop{\longrightarrow}_{\rm free}
  F_1^2 \left[ M^2 - p \cdot p^\prime
               + 2 E_p E_{p}^\prime \right] =
  F_1^2 \left[ M^2 + E_p E_{p}^\prime
               +{\bf p}\cdot{\bf p}^\prime \right]\;.
\label{w00ddfree}
\end{equation}    
Also note, as a consequence of the lower component of the bound-state
spinor $f_{\alpha}(p)$ being substantially smaller than the upper
component $g_{\alpha}(p)$, that $|{\bf p}_{\alpha}|\!\ll\!E_{\alpha}$ 
while $M_{\alpha}\!\simeq\!E_{\alpha}$. This is true even though the 
lower-to-upper ratio $f_{\alpha}/g_{\alpha}$ has been enhanced
considerably in the nuclear medium relative to its free-space value. 
This is an important step towards isolating an observable sensitive 
to the spectral function. Indeed, if the longitudinal component of 
the hadronic tensor is computed in parallel 
(${\bf \hat{p}}^{\prime}\!=\!{\bf\hat{q}}$) kinematics, 
Eqs.~(\ref{w00dd}) and~(\ref{w00ddfree}) reduce to the following
simple expressions:
\begin{mathletters}
\begin{eqnarray}
   W^{00}_{\rm DD}  &=& 
   F_1^2 (E_{p}^\prime+M) 
  \left[\frac{\pi}{p^{2}}g_{\alpha}^{2}(p)\right]
  \left[1 \pm \left(\frac{f_{\alpha}(p)}{g_{\alpha}(p)}\right)
              \left(\frac{|{\bf p}^{\prime}|}{E_{p}^\prime+M}\right)
  \right]^{2}\;, \\
  \label{w00ddpar}
   W^{00}_{\rm DD}\Big|_{\rm free} &=&
   F_1^2 (E_{p}^\prime+M) 
  \left[\frac{1}{2}(E_p\!+\!M)\right]
  \left[1 \pm \left(\frac{|{\bf p}|}{E_p+M}\right)
              \left(\frac{|{\bf p}^{\prime}|}{E_{p}^\prime+M}\right)
  \right]^{2}\;.
\label{w00ddfreepar} 
\end{eqnarray}
\end{mathletters}
The $\pm$ sign in the above expressions corresponds to a missing
momentum ${\bf p}$ either parallel or antiparallel to ${\bf p}'$. 
We observe that up to second-order corrections in the small 
(lower-to-upper) ratios, the hadronic tensor is proportional to 
the energy-like (or mass-like) quantity given in Eqs~(\ref{epm}). 
Yet this energy-like quantity $E_{\alpha}$ is nothing but the 
Fourier transform of the bound-state nucleon density. Thus we 
conclude that, in a mean-field treatment, the nucleon spectral 
function is proportional to the longitudinal response. That is, 
$S(E,{\bf p})\!\propto\!W^{00}_{\rm DD}\!\propto\!E_{\alpha}$. 
Thus, the (Dirac-Dirac component of the) longitudinal hadronic 
tensor is, up to second-order corrections in the lower-to-upper 
ratios, proportional to the nucleon spectral function. Indeed, 
the nucleon momentum distribution may now be easily extracted 
from the longitudinal response. It becomes
\begin{equation}
 \rho_{2}=\;2\;(2j+1)\;(E_{p}+M)
 \left({W^{00}_{\rm DD}}/
  {W^{00}_{\rm DD}\Big|_{\rm free}}\right)\;.
 \label{w00DDratio}
\end{equation}
The momentum distribution for ${}^{4}$He is displayed in
Fig.~\ref{fig3} using various methods for its extraction.  The solid
line gives the ``canonical'' momentum distribution, obtained from the
Fourier transform of the $1S^{1/2}$ proton wave-function [see
Eq.~(\ref{epm})]. The momentum distribution extracted from the
longitudinal response as defined in Eq.~(\ref{w00DDratio}) (dot-dashed
line) is practically indistinguishable from the canonical momentum
distribution.  While it appears that a suitable observable has been
found from which to extract the nucleon momentum distribution, it may
be argued, and justifiably so, that $W^{00}_{DD}$ is not a physical
observable (as $F_{2}$ has been neglected). Hence, the merit of such
an extraction may be put into question. To show that the above
procedure is still robust, we display in the figure (with a dotted
line) the momentum distribution extracted from the full longitudinal
response, namely, one that also includes the anomalous component of
the current. This result remains indistinguishable from the canonical
momentum distribution. Although this behavior is general, it is most
easily understood by limiting the discussion to the case of parallel
kinematics. In this case the longitudinal response becomes equal
to~\cite{gp94}
\begin{equation}
 R_{\rm L} \equiv W^{00} = (E_{p}^\prime+M) 
  \left[\frac{\pi}{p^{2}}g_{\alpha}^{2}(p)\right]
  \left[\left(F_{1}-\xi_{p}^{\prime}\bar{q}
	      F_{2}\right)
    \pm \left(\xi_{p}^{\prime}F_{1} +
                     \bar{q}F_{2}\right)
        \left(\frac{f_{\alpha}(p)}{g_{\alpha}(p)}\right)
  \right]^{2}\;.
 \label{w00full}
\end{equation}
The contribution from the anomalous form factor $F_{2}$ to
the longitudinal response is small because it appears 
multiplied by two out of three``small'' quantities in the
problem: the lower-to-upper ratio, 
$\xi_{p}^{\prime}\!\equiv\!|{\bf p}^{\prime}|/(E_{p}^\prime+M)$,
and $\bar{q}\!\equiv\!|{\bf q}|/2M$. Thus, up to second order 
corrections in these small quantities, the longitudinal response
is given by
\begin{equation}
 R_{\rm L}
  \simeq F_1^2 (E_{p}^\prime+M) 
  \left[\frac{\pi}{p^{2}}g_{\alpha}^{2}(p)\right]
  \simeq F_1^2 (E_{p}^\prime+M) E_{\alpha} \;.
 \label{w00approx}
\end{equation}
The last calculation displayed in Fig.~\ref{fig3} corresponds
to a momentum distribution extracted from the factorization 
approximation using the $cc2$ form for the electromagnetic 
current (long dashed line). The momentum distribution extracted 
in this manner overestimates the canonical momentum distribution 
over the whole range of missing momenta and integrates to 2.9 
rather than 2; this represents a discrepancy of 45 percent.

In summary, the longitudinal response appears to be a robust
observable from which to extract the nucleon momentum
distribution. Experimentally, one should proceed as follows: 
perform a Rosenbluth separation of the $(e,e'p)$ cross section 
so that the longitudinal response ($R_{\rm L}\!\equiv\!W^{00}$) 
may be extracted. This expression should then be divided by the 
corresponding single-nucleon response. Up to a simple and 
unambiguous kinematical factor this yield, at least in the 
plane-wave limit, the nucleon momentum distribution:
\begin{equation}
 \rho_{2}=2\,(2j+1)\,(E_{p}+M)
 \left(\frac{R_{\rm L}}{R_{\rm L}^{\rm free}}\right)\;.
 \label{w00ratio}
\end{equation}
Note that up to second order corrections in various small 
quantities, this form is independent of the small components 
of the Dirac spinors and also of the negative-energy states. 
Moreover, it is also free of off-shell ambiguities. Indeed, 
we could have used the $cc1$ form of the electromagnetic current 
and the results would have remained unchanged. We regard the 
outlined procedure as much more robust than the conventional 
one given in Eq.~(\ref{speca}) because the transverse component
of the hadronic tensor is strongly dependent on the small
components of the wave-function and also sensitive to 
off-shell extrapolations~\cite{gp94}.

\begin{figure}
\bigskip
\centerline{
  \psfig{figure=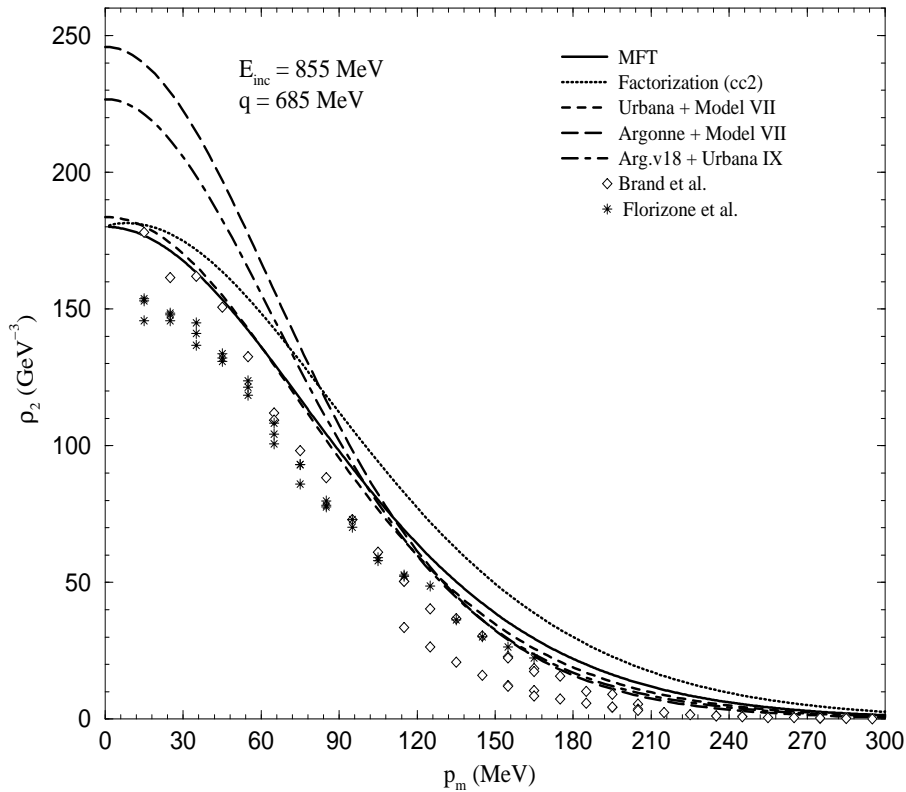,height=5.0in,width=5.5in,angle=-90}} 
 \vskip 0.1in 
 \caption{A comparison between our relativistic calculations, 
	  nonrelativistic calculations reported elsewhere,
          and experimental data for the proton momentum 
	  distribution in ${}^{4}$He. The solid line is our 
	  mean-field calculation while the dotted curve is our
          calculation using the factorization approximation 
	  at incident photon energy of $E_{\rm inc}=855$~MeV 
	  and a momentum transfer of $q=685$~MeV. The 
	  nonrelativistic calculations of Schiavilla {\it et al.,} 
	  are included for both the Urbana (dashed) and the 
	  Argonne (long dashed) potentials as well as the 
	  calculations of Wiringa {\it et al.,} (dashed-dotted). 
	  The NIKHEF data of van den Brand {\it et al.,} for two
          different kinematical settings as well as preliminary 
	  data of Florizone {\it et al.,} ($A1$ collaboration) 
	  which were measured at MAMI (Mainz) are also shown.}
\label{fig4}
\end{figure}

In Fig.~\ref{fig4} a comparison is made between our results and 
nonrelativistic state-of-the-art calculations of the momentum
distribution of ${}^{4}$He. The solid line displays, exactly 
as in Fig.~\ref{fig3}, the canonical momentum distribution. 
We see no need to include the momentum distribution extracted 
from the longitudinal response [Eq.~(\ref{w00ratio})] as it 
has been shown to give identical results.
In addition to our own calculation, we have also included the
variational results of Schiavilla and collaborators~\cite{sch86}, for
both the Urbana~\cite{Lag81} (dashed line) and the
Argonne~\cite{Wir84} (long-dashed line) potentials, with both of them
using Model VII for the three-nucleon interaction. The variational
calculation of Wiringa and collaborators~\cite{Wir91,Wir01,forest96}
(dashed-dotted) has also been included; this uses the Argonne v18
potential~\cite{Wir95} supplemented with the Urbana IX three-nucleon
interaction~\cite{Pub95}. Figure~\ref{fig4} also shows NIKHEF data by
van den Brand and collaborators~\cite{brand91,brand88} as well as
preliminary data from MAINZ by Florizone and
collaborators~\cite{Flor98,Flor01} for three different kinematical
settings. (Results in final form will be submitted shortly.)
Comparisons to the preliminary Mainz data of Kozlov and
collaborators~\cite{koz00,koz01a,koz01b} have also been made (although
the data is not shown). These measurements are consistent, in the
region where comparisons are possible, to the experimental data of
both van den Brand and Florizone. Thus, high-quality data for the
momentum distribution of ${}^{4}$He is now available up to a missing
momentum of about 200~MeV. We find the results of Fig.~\ref{fig4}
quite remarkable. It appears that a simple relativistic mean-field
calculation of the momentum distribution rivals --- and in some cases
surpasses --- some of the most sophisticated nonrelativistic
predictions. The mean-field calculations reported here, with the
scalar mass adjusted to reproduce the root-mean-square charge radius 
of ${}^{4}$He, provide a good description of the experimental data.
Still, theoretical predictions of the momentum distribution
overestimate the experimental data by up to 50-60\%. Part of the
discrepancy is attributed to distortion effects which are estimated
at about 12\%~\cite{Flor98,sch90}. However, distortions are not able
to account for the full discrepancy. We have argued earlier that an
additional source of error may arise from the factorization
approximation [see Eq.~(\ref{speca})] used to extract the spectral
function from the experimental cross section.  The use of an off-shell
prescription, such as the $cc1$ prescription for $\sigma_{eN}$,
combined with the in-medium changes in the lower-component of the
Dirac spinors contaminate the extraction of the spectral function. One
could estimate the source of the off-shell ambiguity by monitoring the
variations in the spectral function as other on-shell equivalent forms
for the single-nucleon current are used. While such an approach is
useful for estimating a theoretical error, it is clearly not
sufficient to eliminate it. We are confident that the approach
suggested here, based on the extraction of the spectral function from
the longitudinal response, is robust.  While the method adds further
experimental demands, as a Rosenbluth separation of the cross section
is now required, the extracted spectral function appears to be weakly
dependent on off-shell extrapolations and relativistic effects. If
deviations between experiment and theory still persist, these may
suggest physics beyond the baseline model, such as violations to the
impulse approximation or to the independent particle picture.

\section{Conclusions}
\label{sec:concl}

To summarize, we have calculated the spectral function of ${}^{4}$He
in a plane-wave approximation to the $(e,e'p)$ reaction using a fully
relativistic formalism. We have taken advantage of an algebraic trick
originally introduced by Gardner and Piekarewicz and of our recently
developed relativistic formalism for quasifree processes to arrive at 
transparent, analytical results for the quasifree reaction. We have 
found that a simple relativistic mean-field calculation of the momentum 
distribution in ${}^{4}$He rivals --- and in some cases surpasses --- 
some of the most sophisticated nonrelativistic predictions to date.
These calculations attempt to provide theoretical support to the
recently measured, but not yet fully analyzed, $A1$ collaboration data
from Mainz. The final experimental reports are expected to be published
shortly.

We have also demonstrated that a more robust procedure, relative to
the conventional factorization prescription, exists for extracting the
spectral function. This procedure uses the ratio of quasifree to
single-nucleon longitudinal responses, rather than the ratio of cross
sections, to isolate the momentum distribution. We have shown that the
longitudinal ratio is fairly insensitive to off-shell ambiguities and
to the negative-energy part of the spectrum, as both of these effects
appear as second-order corrections to a ``canonical'' momentum
distribution. This ceases to be true in the case of the ratio of cross
sections because the transverse response is sensitive to both effects.
While this procedure relies on a Rosenbluth (L/T) separation
of the quasifree cross section, and thus presents the experimentalist
with a more demanding task, the experimental field has evolved to such
a level of maturity that L/T separations are now almost routine.
Indeed, in a recent publication~\cite{goff97} a Rosenbluth separation
of the ${}^{3}{\rm He}(e,e'p)$ cross sections was made in order to
extract ``longitudinal'' and ``transverse'' spectral functions in the
hope of resolving the anomaly in the longitudinal-transverse ratio
alluded to in the introduction. We speculate that the sensitivity 
of the transverse response to more complicated dynamical processes
might be partially responsible for the quenching of the 
longitudinal--transverse ratio.

Finally, although in this article we focused exclusively on the
spectral function, the formalism presented here may be extended in a
straightforward fashion to the calculation of spin observables in
quasifree electroproduction processes. Indeed, we speculate that,
because the ratio of quasifree cross sections are fairly insensitive
to distortion effects, spin-observables may be a more fruitful testing
ground for our relativistic plane-wave model. Moreover, our formalism
may be easily extended to neutrino-induced reactions. It has been
suggested that a measurement of the ratio of neutral to
charge-changing neutrino-nucleon scattering may provide a clean
signature of the strange-quark content of the nucleon~\cite{cjh01}.
This measurement is believed to be free from most of the
uncertainties, such as radiative corrections, that hinder the
parity-violating electron scattering program. Yet neutrino experiments
suffer from very low counting rates. To remedy this situation neutrino
experiments employ large quantities of nuclear targets (such as
organic scintillators) that provide both the target and the detection
medium. Thus neutrinos interact, not only with the free protons in the
target, but also with protons and neutrons bound to nuclei; hence, one
must compute quasifree $(\nu,\nu^{\prime}\,p)$ and $(\nu,\mu^{-}p)$
cross sections. (Of course, one must integrate the quasifree cross
section over the undetected outgoing neutrino). Therefore, the
relativistic plane-wave formalism presented here is ideally suited,
after including an additional axial-vector term in the single-nucleon
current, to predict ratios of quasifree neutrino-nucleus cross
sections in the quasifree region.

\acknowledgments
We are grateful to Drs. S. Gilad, A. Kozlov, A.J. Sarty, and
the $A1$ collaboration as well as Dr. J.F.J. van den Brand 
for providing us with their data for this process and for 
many helpful discussions on this subject. We would also like
to thank Drs. R. Schiavilla and R. B. Wiringa for providing 
us with their theoretical calculations. This work was supported 
in part by the United States Department of Energy under Contract
No. DE-FG05-92ER40750 and in part by a joint fellowship from the 
Japan Society for the Promotion of Science and the United States 
National Science Foundation.  

\vfill\eject
\vfill\eject
\vfill\eject
\vfill\eject
\end{document}